# NOVEL HYPOSTASIS OF OLD MATERIALS IN OXIDE ELECTRONICS: METAL OXIDES FOR RESISTIVE RANDOM ACCESS MEMORY APPLICATIONS

*A. Pergament[∗], G. Stefanovich, A. Velichko,*
*V. Putrolainen, T. Kundozerova and T. Stefanovich*

Physics and Technology Department, Petrozavodsk State University,
185910, Petrozavodsk, Russia

## ABSTRACT

Transition-metal oxide films, demonstrating the effects of both threshold and nonvolatile memory resistive switching, have been recently proposed as candidate materials for storage-class memory. In this work we describe some experimental results on threshold switching in a number of various transition metal (V, Ti, Fe, Nb, Mo, W, Hf, Zr, Mn, Y, and Ta) oxide films obtained by anodic oxidation. Then, the results concerning the effects of bistable resistive switching in MOM and MOS structures on the basis of such oxides as $V_2O_5$, $Nb_2O_5$, and NiO are presented. It is shown that sandwich structures on the basis of the $Au/V_2O_5/SiO_2/Si$, $Nb/Nb_2O_5/Au$, and Pt/NiO/Pt can be used as memory elements for ReRAM applications. Finally, model approximations are developed in order to describe theoretically the effect of nonvolatile unipolar switching in Pt-NiO-Pt structures.

## 1. INTRODUCTION

### 1.1. Contemporary Trends in Oxide Electronics

Metal oxides (transition metal oxides, TMO, included) have been known for a long time, and their properties would seem to have been studied in detail and quite comprehensively. While the utility of metal oxide films, e.g., as good insulators has been well known, their active properties have been, so to say, in the shadow just till recently. One of the most pronounced among those active properties is the effect of electronic switching [1].

At present, the problem of standard Si-based CMOS technology restrictions (first of all, connected with a limit for further scaling) are intensively discussed in the literature [2-4].

[∗] E-mail: aperg@psu.karelia.ru

Alternative approaches are based on either a different physical mechanism or even a drastic change of computing paradigms (quantum computer or neuroprocessors). The first approach, closest to existing circuits, uses novel materials and devices [3]. One of such devices, termed as a Mott transition field-effect transistor, or MTFET, has been proposed in [5]. Approaches based on new physical principles are well known. They are, for example, spintronics, superconducting (SC) electronics, single-electronics, molecular electronics, etc [3, 4]. One of these new directions, oxide electronics [6-9]; the idea of using the unique properties and physical phenomena in TMO, underlies such an approach.

A set of valence states, associated with the existence of unfilled $d$-shells in the atoms of transition metals, leads to formation of several oxide phases with different properties, ranging from metallic to insulating. On the other hand, it is specifically the behavior of $d$-electrons in the compounds of transition metals that is responsible for the unique properties of these materials, causing strong electron-electron correlations, which play an important role in the mechanisms of such phenomena as, e.g., metal-insulator transitions (MIT), high-$T_c$ SC, and colossal magnetoresistance, which are inherent to many TMOs [8-10].

Transition-metal oxide films, demonstrating the effect of nonvolatile resistive switching, have been recently proposed as candidate materials for storage-class memory [11, 12]. On the basis of $I$–$V$ characteristics, the switching behaviors can be classified into two types: unipolar (nonpolar) and bipolar [12].

Conceptually unsophisticated design of such a MOM memory cell allows easily scalable cross-point ReRAM architecture with nanometer cell dimensions. This, in turn, provides the basis for 3D integrated terabit memory with a multi-layer stackable structure [13].

In this work we report the results on the effects of bistable resistive switching in MOM and MOS structures on the basis of some transition metal oxides. However, first we briefly describe some our previous results on threshold switching in a number of various TMOs in order to somehow elucidate rather more the following problem. A natural question arising is: why in some cases, one encounters threshold switching with no memory effects, whereas in other cases (frequently, on those same materials!) memory switching is really revealed. Finally, we present some model approximations concerning nonvolatile unipolar switching in Pt-NiO-Pt structures, with the nickel oxide merely as an example.

### 1.2. Electrical Switching in Metal Oxides

Electrical switching of a kind can be observed in a great variety of materials in many different forms and structures [1]. Current-voltage characteristics of such systems are usually classified into two categories: those with S-type and with N-type (sometimes termed as "Z-type") NDR – see Figure 1. N-type switching is also often referred to as "VCNR" (voltage-controlled negative resistance), and S-type – as "CCNR", i.e. current-controlled negative resistance [14]. As a rule, these phenomena are observed in amorphous insulators and semiconductors [14-17]. Electronic switching in amorphous thin films can be divided into two general types: (a) monostable threshold switching, in which continuous electrical power is required to maintain the highly conducting ON state; and (b) bistable memory switching, in which both ON and OFF states can be maintained without electrical power.

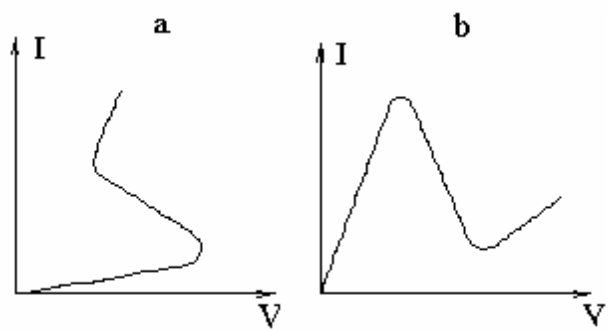

Figure 1. I-V curves with (a) S-type and (b) N-type NDR.

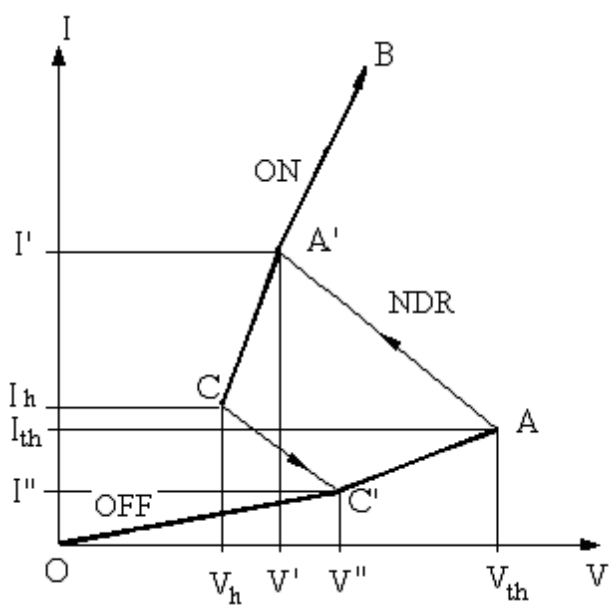

Figure 2. Schematic current-voltage characteristic (O-C′-A-A′-B-A′-C-C′-O) of a switching device with the S-type negative differential resistance region. The O-C′-A branch represents the OFF state, C-A′-B – ON state and A-A′ and C-C′ - NDR regions. $V_{th}$ and $V_h$ ($I_{th}$ and $I_h$) are the threshold and holding voltages (currents), respectively.

In other words, for threshold switching, the OFF-to-ON transition is absolutely reversible and repetitive, whereas for memory switching, a large voltage pulse is usually applied in the erase operation cycle to switch a device from the ON back to OFF state.

The switching phenomena are usually described in terms of ON and OFF states (Figure 2), but switching devices are not always rigidly bistable in their operation, i.e. they are not always "digital" devices. In some cases, a continuous range of intermediate states is observed between the ON and OFF states, giving an "analogue" memory effect (Figure 3).

It should be noted that there is little difference between the "S-shaped" (Figure 3, a) and "threshold" (Figure 3, b) I-V curves. Therefore, we will not differentiate I-V characteristics in this sense, and the terms "S-type switching", "S-NDR", etc will be used only to indicate the type of NR, i.e. to distinguish it from the N-type NDR (see Figure 1, a and b).

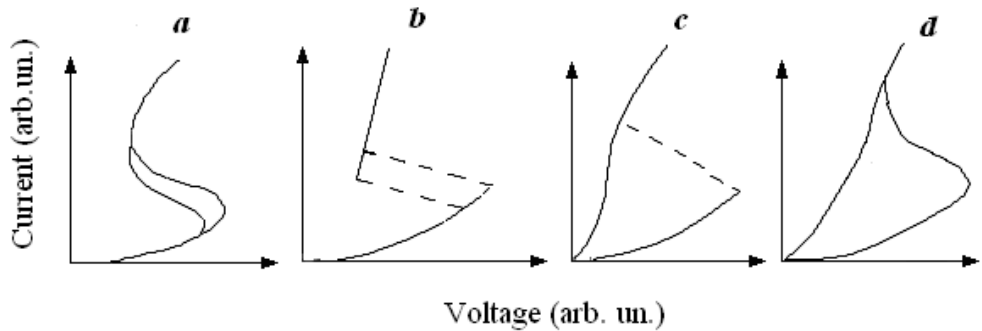

Figure 3. S-shaped (a), threshold (b) and memory (c, d) I-V curves; bistable (c) and "analogue" (d) memory switching [17].

Originally, the switching effect with current-controlled negative resistance was found and investigated in chalcogenide-glass semiconductors (CGS) [15-17]. Current-voltage characteristics of some CGS-based thin-film sandwich structures after *electrical forming* (the process of electroforming will be discussed later) are S-shaped. Similar phenomena have also been observed in oxides (mainly those of transition metals) and oxide glasses, amorphous and polycrystalline Si and other semiconductors, halides, nitrides, sulfides of metals, carbon-containing materials, particularly, nano-tubes, organic compounds, such as conducting polymers and self-assembled monolayers, and many other materials [1, 12, 14-18].

As-fabricated devices rarely show threshold or memory switching effects without an initial modification of their structure, a process which is usually called "forming" or electroforming (EF). There is a striking similarity in the electroforming behavior observed in a wide range of oxide, halide, sulfide, polymer, and CGS films, and further we will follow the discussion of this problem in review [15].

Electroforming is achieved by the application of suitable voltage pulses which are always higher in magnitude than the subsequent operating pulses. This invariably produces an irreversible change in the characteristics of a device, often with a substantial decrease in the overall terminal resistance. The processes involved in electroforming depend on the nature and quality of the thin film, the geometrical structure of the device and, in some cases, the electrode material.

The changes can be either structural, involving a movement of material from one part of the device to another, or they can be electronic, wherein a quasipermanent change in the occupancy of some electronic states takes place. The changes can occur throughout the bulk of the film or in a localized region. The most common localized effect is the formation of a filament of highly conducting material – a switching channel. Such permanent filament formation is a consequence of temporary filamentary breakdown, often observed in MIM or MSM sandwich structures [15]. However, electroforming differs from dielectric breakdown in the sense that it is a non-destructive process (or, in other words, it is an irreversible metal-insulator transition [19]), i.e., it rather resembles a self-healing type of breakdown [20].

It should be emphasized that the EF characteristics of different materials are generally ill defined and exhibit large variations, even for devices fabricated under the same conditions.

Nevertheless, since this process is observed in a wide variety of insulating films, it is unlikely to result from a peculiarity of any one system. In the case of amorphous thin film structures, switching channel formation can occur via crystallization of an amorphous film, stoichiometric changes, diffusion of the electrode material into the film, or ionization of deep traps [15, 17].

All of these changes usually refer to a localized modification of the structure in the area of the filament. The modification of the initial structure during the forming process is the most important factor determining the subsequent switching operation. This is because, in most cases, forming creates a "new device" within the original structure, the characteristics of which determine the ensuing threshold or memory switching operation.

Note, that in both threshold-type devices and those of the memory type, the initial switching mechanism appears to be the same [15]. It is initiated by field-dependent non-ohmic conductivity and a consequent instability. Whether what follows is threshold switching, memory switching or, in some cases, destructive electrical breakdown depends upon the properties of the material and on the presence or absence of suitable feedback in the system. For *memory switching* the active material must be capable of changing in some way (e.g. an overall or localized change in the electronic, atomic or microscopic structure) into a permanent conducting state, but one that can be reversed to the OFF state by a suitable current (energy) pulse.

Obviously, the system must also be able to absorb the reversing pulse without destructive breakdown. The suggested models for memory switching can therefore be divided into two broad categories: electronic or structural [15]. The former is based on the long term storage of charge (i.e. without any major structural modification) to account for the non-volatile nature of the switch. One of the most commonly proposed charge storage sites are traps, either in the bulk or at an interface between two dissimilar materials. The necessary characteristic of such traps is that they have a release time comparable to the retention time of the memory which may range from few hours to many years [12, 15-17].

In CGS devices, memory switching depends upon a structural rearrangement [15, 17]. During the initial forming pulses a filament of material crystallizes in the amorphous structure; this normally happens near the center of the device where the highest temperatures are attained. This modified material has a higher conductivity than the bulk and therefore becomes the preferred current path during subsequent voltage pulses. In the erase operation (i.e. to switch from the ON to OFF state) large voltage pulses with steep trailing edges are applied, melting the crystalline filament and causing it to solidify in an amorphous form.

To write in again (i.e. to switch from the OFF to ON state), pulses with more gradual trailing edges are used, also melting the filament but allowing the material to solidify in a crystalline form [15]. Recently these materials have been "reinvented" and proposed as candidate materials for the so-called phase-change memory (PCM) [21]. The model takes into account the dependence of the carrier density on the electric field strength [15, 17, 22-24].

We will return to the discussion of the switching mechanisms in the subsequent sections. Some relevant references concerning this problem, the problem of oxide-based ReRAM (as well as a more complete list of the materials showing VCNR and CCNR), can also be found in recent original articles, reviews and monographs [1, 14-18, 25, 26].

## 2. THRESHOLD SWITCHING IN ANODIC OXIDE FILMS

### 2.1. Sample Preparation and Experimental Procedures

Oxide films under study were prepared by oxidation of corresponding metals in electrolytes. Electrochemical (anodic) oxidation permits the synthesis of high quality homogeneous dielectric oxide films [27-29]. The thickness d of the films can be easily controlled from approximately 0.05 to 0.5 µm by the anodizing voltage V: $d \approx \alpha V$, where $\alpha$ = const. For example, $\alpha$ = 1.6 nm/V for Ta and 2.2 nm/V for Nb. For a long time, these films have been used in technical applications just as dielectrics in electrolytic capacitors (Ta, Nb, Al). There have been also some attempts to use $Ta_2O_5$, $ZrO_2$, $HfO_2$ and some other oxide films as high-*k* gate insulators in MIS devices. However, anodic oxides of transition metals are of considerable interest not only due to their wide technical applications as dielectrics in electrolytic capacitors or electrochromic materials (W, Mo, V), or as corrosion-resistant and decorative coatings. In addition, since the anodic oxide films (AOF) are usually amorphous, they are receiving increased attention because of general interest in the disordered state, especially in the problem of an interplay between disorder and electron interactions in the strongly correlated systems, such as materials with metal-insulator transitions [29, 30].

Contrary to many transition metals, anodic behavior of vanadium in electrolytes is rather unusual. Anodic films on vanadium can be formed only in exceptional cases, including electrolytes based on acetic acid or acetone [28]. When anodizing vanadium, the main problem is that the AOF is rapidly undergoing dissolution in aqueous electrolytes [28-30].

Taking the aforementioned into account, for vanadium anodization, the electrolyte based on either acetone or acetic acid was used. Other metals (Ti, Fe, Nb, Mo, W, Hf, Zr, Mn, Y, and Ta) were oxidized in standard condition [1, 27, 28, 31]. Current-voltage (I-V) characteristics were recorded by the two-probe method with a Keithley 2410 Sourcemeter and a probe station SUSS PM5. The voltage scan rate was 10–25 V/s, and the compliance current was varied in the range from 2 to 500 µA. In some cases, a standard and simple to operate oscilloscopic technique [31] was also used for mass measurements.

### 2.2. Electroforming and Switching

Initial structures demonstrate the nonlinear and slightly asymmetric current-voltage characteristics. The resistance at zero bias, measured with the point contact, is in the range $10^7$-$10^8$ $\Omega$ for the vanadium anodic oxide, which is the most highly conductive among all the materials studied. When the amplitude of the applied voltage reaches the forming voltage $V_f$, a sharp and irreversible increase in conductivity is observed and the I(V) curve becomes S-shaped. With increasing current, the I-V characteristic may change until the parameters of the switching structure are finally stabilized. The process outlined above is qualitatively similar to the electroforming of the switching devices based on amorphous semiconductors [16, 17].

Thus, the first stage of the forming does not differ from the conventional electrical breakdown of oxide films. However, if the post-breakdown current is limited, this results in formation of a switching channel rather than a breakdown one. The latter would be expected to have metallic-like conductivity and no negative resistance in the I-V characteristic. It is quite evident that the phase composition of this switching channel must differ from the material of the initial oxide film, because the channel conductivity exceeds that of an unformed structure by several orders of magnitude. Forming does not always result in an S-type characteristic. In some instances a transition of the structure to a high conductivity state (R~100 Ω) with ohmic behavior takes place; i.e. in this case breakdown rather than forming occurs.

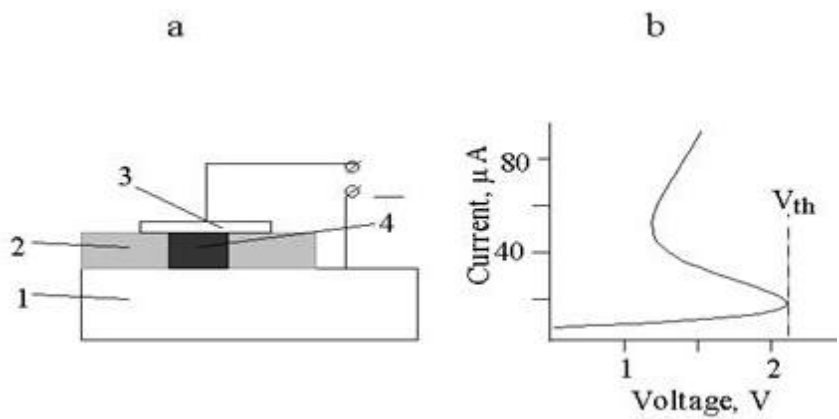

Figure 4. (a) The sandwich switching structure (schematic): 1 – vanadium, 2 – anodic oxide film, 3 – Au electrical contact, 4 – switching channel (VO$_2$, as will be shown below); (b) the I-V characteristic for one of the samples at room temperature after EF.

The voltage-current characteristic for electroformed vanadium-AOF-metal structure is shown in Figure 4. The parameters of the structures (threshold voltage $V_{th}$ and current $I_{th}$, resistances of OFF and ON states) may vary by up to an order of magnitude from point to point for the same specimen. Such a wide range of variation of the $V_{th}$ and $R_{off}$ values, as well as the absence of a correlation between these switching parameters and the parameters of the sandwich structures (the electrode material and area, the film thickness), leads to the conclusion that resistance and threshold parameters are mainly determined by the forming process. Conditions of the EF process cannot be unified in principle, because the first stage of forming is linked to breakdown, which is statistical in nature.

As a result, the diameter and phase composition of the channel (and, consequently, its effective specific conductivity) vary with position in the sample. This accounts for the scatter in the parameters and for the seeming absence of their thickness dependence.

Similar behavior has been observed for the other materials. On average, for d ~ 100 nm, a typical value of $V_{th}$ is 1-5 V; i.e., the threshold field is $(1-5) \cdot 10^5$ V/cm at room temperature. The greatest differences between these oxides are observed when the temperature dependences of their threshold parameters, such as $V_{th}$, are measured.

Figure 5 shows the $V_{th}(T)$ curves for the sandwich structures based on iron, vanadium, titanium and niobium. For Ta$_2$O$_5$, HfO$_2$ and MoO$_3$, no reproducible results were obtained because of instabilities of their voltage-current characteristics with temperature. As the temperature increased, $V_{th}$ decreased, tending to zero at some finite temperature $T_o$. This is best exhibited by the vanadium and titanium oxide structures.

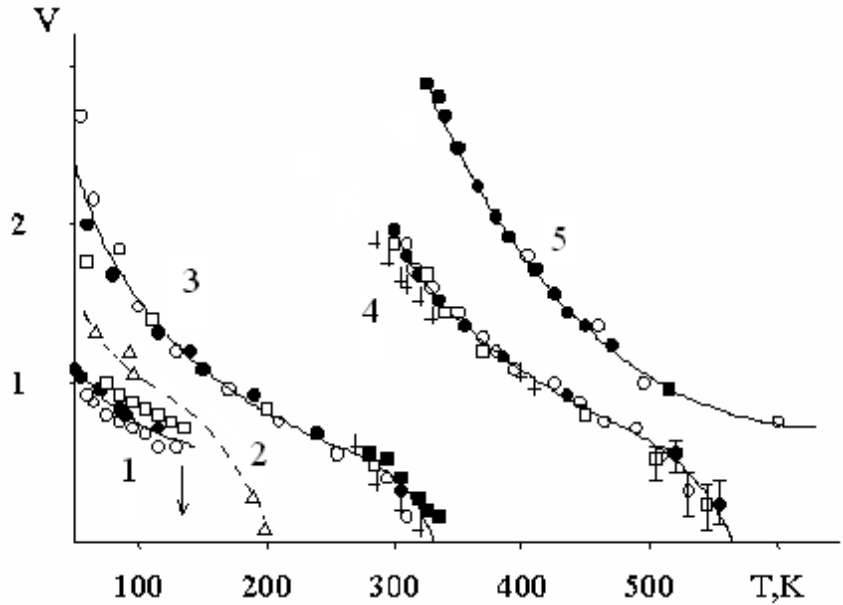

Figure 5. Threshold voltage (in arbitrary units $V=V_{th}(T)/V_{th}(T')$) for the MOM structures on the basis of different oxides as a function of ambient temperature. 1 – Fe, 2 – W, 3 – V, 4 – Ti, and 5 – Nb anodic oxide. Data for different samples have been averaged and normalized to a certain temperature T' (different for different materials).

The values of $T_o$ vary considerably for different oxides, but they are nearly identical for different structures of the same oxide. S-type switching has not been observed in Zr, Y, and Mn oxide films. Instead, these materials have sometimes demonstrated VCNR at T=293 K (Zr and Y AOFs, and Al oxide as well) or at T=77 K (Mn oxide), although the NR region was "degenerate" in some cases. The process of EF for these materials was hindered, likewise that for Ta, W, Hf, and Mo.

Next we consider the switching behavior of vanadium-based sandwich devices in more detail. Some of the previously obtained experimental results [24] are presented in Figure 6 and can be summarized as follows.

On the basis of the study of I-V characteristics in a wide temperature range (Figure 6, *a*), it has been shown that, as the field strength increases, the transition temperature (i.e. the temperature at which the switching event inside the channel occurs) decreases (Figure 6, *b*). In addition, numerical simulation of the free charge carrier density has been made (Figure 7) using the relation $\sigma = ne\mu$, where $\sigma$ is the channel conductivity measured experimentally, and $\mu = 0.5$ cm$^2$ V$^{-1}$ s$^{-1}$ [32] is the mobility of electrons in vanadium dioxide.

One can see that, at $T_0 < 200$ K ($E > 10^5$ V cm$^{-1}$), the value of *n* becomes less than the critical Mott density $n_c$ in equation for the Mott criterion [32], which for vanadium dioxide has been estimated to be $n_c \sim 3 \cdot 10^{18}$ cm$^{-3}$ [1, 24]. Note that this behavior of $n(T)$ in the low-temperature (LT) region can not be explained by a dependence of $\mu$ on temperature – see curve 2 in Figure 7.

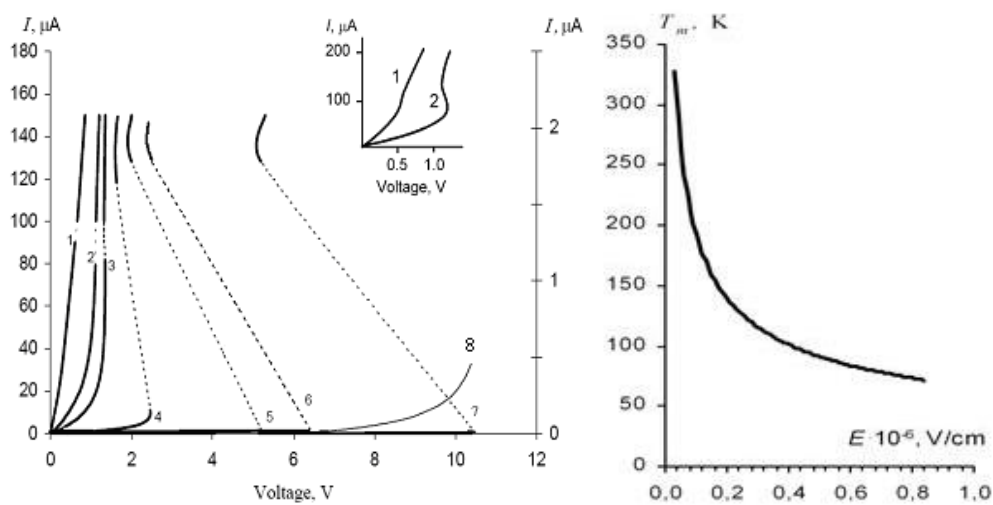

Figure 6. (*a*) Experimental current–voltage characteristics of the VO$_2$ based switch at various ambient temperatures $T_0$ (K): (1) 293; (2) 241; (3) 211; (4) 144; (5) 91; (6) 70; (7) 15. Curve (8) represents the OFF state for $T_0 = 15$ K (right-hand current axis; that is it is the same as curve 7, but in an exaggerated scale), and the insert shows curves (1) and (2) also on scale up. Dashed lines indicate unstable NDR regions of the I-V curves. (*b*) Field dependence of the critical switching temperature $T_m$. As the field $E \to 0$, $T_m$ tends to the equilibrium value of the MIT temperature, which is equal to $T_{t0} = 340$ K for VO$_2$ [24].

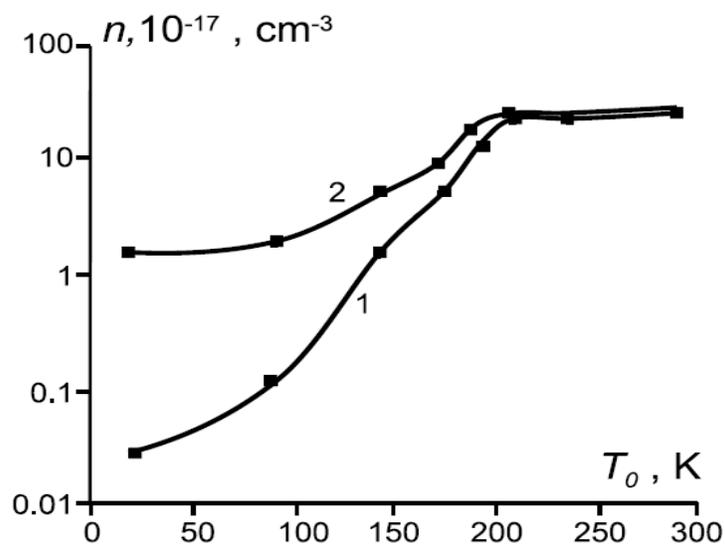

Figure 7. Maximum electron density in the switching channel as a function of ambient temperature, for different I-V curves in Figure 6, *a*, calculated on the basis of numerical modelling [24]. The calculations were made for the cases of $\mu = $ const (curve 1) and $\mu = \mu(T)$ (curve 2), where $\mu(T) \sim \exp(-W_\mu/kT)$ is a typical temperature dependence for the mobility in strongly correlated systems [1, 24, 32].

## 2.3. Switching Mechanism

As was discussed above (in Section 1.2) the switching mechanism has for a long time been a matter of especial concern. Many a one is likely to remember heated discussions of the 1970s about the mechanism of electrical switching in different systems, CGS included, and the principal issue was: whether the mechanism is primarily thermal or electronic [16]. There have also been a lot of works treating the switching effect as a metal-insulator transition occurring in electric field [16, 31, 33-41]. The field effect upon the MIT in $VO_2$ has been studied previously, both theoretically and experimentally, in a number of works [1, 4, 24, 33-36]. Particularly, a thermodynamic analysis based on the standard phenomenological approach [4, 34], using the equation for the free energy, shows that the shift of the transition temperature in electric field is

$$\Delta T_t \sim T_t \varepsilon \varepsilon_0 E^2 / q, \tag{1}$$

where $q = 250$ J cm$^{-3}$ is the latent heat of the transition [34]. The change of $T_t$ is negligible in this case (~ 1 K for $E \sim 10^5$ V cm$^{-1}$). Also, since the entropy of $VO_2$ increases at the transition into metallic phase, the value of $T_t$ increases with increasing $E$, i.e. $\Delta T_t > 0$ in equation (1) [4]. A decrease of $T_t$ in an electric field (Figure 6, *b*), and finally its fall down to zero at a certain critical field $E_c$, can be obtained using a microscopic, not thermodynamic, approach based on the detailed MIT mechanism. Unfortunately, as noted in [4], we have no a quantitative theory to describe such a transition. Nevertheless, taking into account the fact that the MIT in $VO_2$ is an electronically-induced transition, we next consider the following model. Let us imagine an array, e.g. three-dimensional lattice, of partly ionized one-electron sites (positively-charged centres) with the localization radius $a_H$ and the number of free electrons $n$, $0 < n < n_c$. This model describes a doped semiconductor or an EI-phase [37] at $n < n_c$, or, e.g., a material exhibiting a temperature-induced MIT, such as $VO_2$, at $T < T_t$. Application of an electric field to this system, taking also into account the current flow (since we consider the switching effect, not a stationary field effect), will result in both the thermal generation of additional carriers due to the Joule heat, and in the field-induced generation. The latter is due to the autoionization caused by the Coulomb barrier lowering – a phenomenon analogous to the Poole-Frenkel effect. When the total concentration of free carriers reaches the value of $n = n_c$, the transition into the metal state (i.e. switching of the structure into the ON-state) occurs. In a material with the temperature-induced MIT, this switching (in a relatively low field) occurs just due to the heating of the switching channel up to $T = T_t$.

There exists however one more possibility: the switching can occur at $n < n_c$ and $T < T_t$, but when $E = E_c$, where $E_c$ is a critical field. This $E_c$ is the field strength for which there are no bound states in the Coulomb potential of positive centres.

It has been shown [24] that the dependence of $E_c$ on electron density $n$ might be written (according to the scaling theory [1]) as:

$$E_c = E_0 (1 - n/n_c)^\gamma, \tag{2}$$

i.e. the switching event in a high field can occur at $E = E_c$ when the condition $n = n_c$ is not fulfilled yet. Here $n_c$ is the critical Mott transition density, which is easy to calculate from the Mott criterion [32]:

$$a_H n_c^{1/3} \approx 0.25, \tag{3}$$

where $a_H = \varepsilon\hbar^2/m^*e^2$ is the effective Bohr radius, $\varepsilon$ – the static dielectric permittivity, and $m^*$ – the effective mass of a charge carrier. In the work [24] we have also obtained a relation between the maximum channel temperature $T_m$ and electron density at the switching point ($E = E_c$):

$$T_m/T_t = \frac{1-(1-n/n_c)^{\gamma/2}}{1-(kT_{t0}/W)\ln(n/n_c)}, \tag{4}$$

where $T_{t0}$ is the equilibrium transition temperature, and $W$ is the conductivity activation energy. The graphs of these dependences are presented in Figure 8, where $\gamma$ and $(kT_{t0}/W)$ are considered as variable parameters.

One can see that the forms and positions of the curves in Figure 8 are almost independent of the variation of the parameter $(kT_{t0}/W)$ in the range 0.05 to 0.30 which corresponds to the activation energy $W \sim 0.15$–$0.9$ eV for $T_{t0} = 340$ K (for vanadium dioxide, $E_g \sim 1$ eV, and $W \sim 0.5$ eV at $T \to T_t$). As to the $\gamma$ exponent, the best accordance with the experimental data is achieved for $\gamma \sim 10$ (*cf.* curves 2 and *d* in Figure 8). This value of $\gamma$ is in a good agreement with experimental data for materials with MIT [1, 24]. It should be noted that, though the coincidence of the experimental data (curves 1 and 2 in Figure 7) and theoretical results (curves *d* in Figure 8) is only approximate, nevertheless it is quite satisfactory.

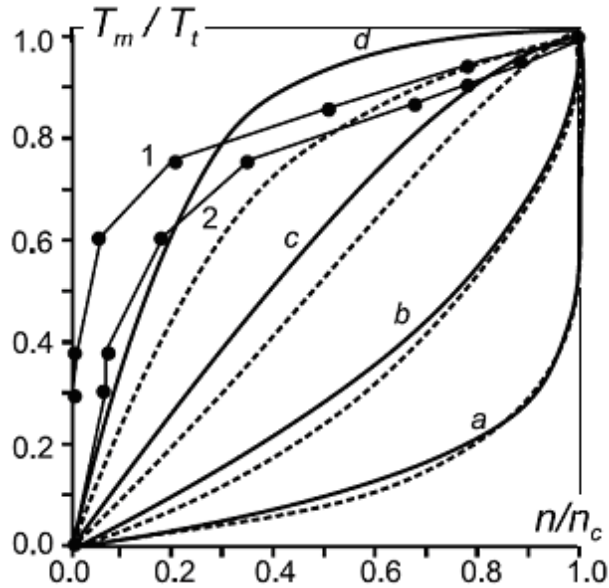

Figure 8. Maximum switching channel temperature as a function of electron density at $E = E_c$ (equation (11)) with $\gamma = 0.3$ (*a*), 1 (*b*), 3 (*c*) and 10 (*d*), and $(kT_t/W) = 0.05$ (solid lines) and 0.3 (dotted lines). Experimental data (curves 1 and 2) correspond to those in figure 7.

In conclusion, the proposed model not only allows the qualitative description of the switching mechanism, but it is in quantitative agreement with the experimental results, in particular, with those concerning the critical concentration (Figures 7 and 8). Thus, the main features of switching both in transition metal compounds and in other strongly correlated systems can be naturally explained within the frameworks of a universal switching mechanism based on the electronically-induced metal-insulator transition. In some cases, a preliminary electroforming is required. The switching channel forms in the initial structure during such electroforming resulted from the electrothermal and electrochemical processes under the action of the applied forming voltage [1, 24, 31].

This channel consists (partly or completely) of a material [31] which can undergo a MIT from one stable state into another at a certain critical temperature $T_t$ or electron density $n_c$. The S- or N-shaped I-V curve is conditioned by the development of an electrothermal instability in the switching channel.

Due to the effect of Joule heating, when the voltage reaches a critical value $V = V_{th}$, the channel is heated up to $T = T_t$ and the structure undergoes a transition from the insulating OFF state to the metallic ON state (for the case of S-switching).

This is the model of "critical temperature" (i.e. a simple electrothermal mechanism, albeit taking into account the specific $\sigma(T)$ dependence of the material at the MIT), though the mechanism of the MIT itself is, of course, essentially electronic. When the ambient temperature $T_0$ is much less than the transition temperature, and the value of $E_{th}$ is high enough, the effect of electronic correlations upon the MIT is feasible. In high electric fields electronic effects influence the MIT so that a field-induced increase in the charge carrier concentration (due to either injection from contacts or impact ionization, or due to field-stimulated donor ionization – i.e. the Poole-Frenkel effect [24]) leads to the elimination of the Mott-Hubbard energy gap at $T < T_t$ [1, 39].

This effect may be treated also as a lowering of $T_t$ due to an excess negative charge (electrons), and the dependence of $V_{th}$ on $T$ deviates from the behavior described by the critical temperature model. In this case, switching commences when a certain critical electron density $n_c$ is achieved. In the equilibrium conditions, the value of $n_c$ is achieved merely due to the thermal generation of carriers at Joule heating up to $T \sim T_t$. Also, in even higher fields, switching occurs at $E = E_c$, when $T_m < T_t$ and $n < n_c$.

To summarize, the results presented above, as well as the other data on the switching in transition metal oxides [38-44], indicate that current instabilities with the S-type NR exhibit several common features. In particular, for each of the investigated materials there is a certain fixed temperature $T_o$, above which switching disappears. At $T < T_o$, the threshold voltage decreases as the temperature rises, tending to zero at $T = T_o$.

Comparison of these temperatures with the temperatures of MIT for some compounds ($T_t$=120 K for $Fe_3O_4$, ~200-250 K for $WO_{3-x}$, 340 K for $VO_2$, ~ 500 K for $Ti_2O_3$, and 1070 K for $NbO_2$) – see Figure 5 – shows that the switching effect is associated with the insulator-metal transition in an electric field. The channels, consisting of these lower oxides, are formed in initial anodic films during preliminary electroforming.

# 3. MEMORY SWITCHING IN HETEROSTRUCTURES BASED ON TRANSITION METAL OXIDES

## 3.1. Bistable Resistive Switching in Au/V$_2$O$_5$/SiO$_2$/Si Junctions

Two-terminal sandwiched structures with a memory switching are conditionally classified into two types according to the mechanism involved: those with an electrical capacity change (capacitance memory) [45] and with a conductivity change (resistance memory) [46]. The latter can contain several layers, including dielectric ones, in order to improve the characteristics of reversible electrical switching. These structures may also change the capacity [47], though this feature is not critical for their operation. Recently, bistable resistance switching has been found in transition metal oxides revived the interest to the phenomenon of reversible electrical switching first observed in various types of disordered chalcogenide-based semiconductors by Ovshinsky [48]. Since then, the effect was observed in Nb$_2$O$_5$, TiO$_2$, NiO, Al$_2$O$_3$, CeO$_2$, various complex oxides with a perovskite structure and many other materials [49-52].

Metallic conducting filaments in the low-resistance state (LRS) of SiO/V$_2$O$_5$ thin films, used as memory elements, have been observed in [53]. Also, the effect of non-volatile memory has been found in Si/VO$_2$/Au structures [54], and the switching mechanism has been attributed to the phenomena in the Si-VO$_2$ heterojunction.

In the work [55], $p+$-Si-SiO$_2$-SnO$_2$-metal structures with a tunnel SiO$_2$ layer have been studied. The physical mechanism of memory and switching is attributed to redistribution of ionized impurity in SnO$_2$ because of electromigration occurring due to the Joule heating of the switching channel. After the voltage is switched off, new impurity distribution becomes frozen and ensures a decrease of the current activation energy and, thereby, an increase of the density of states participating in the tunneling transitions.

In this section we report on the resistance switching characteristics in vanadium oxide based structures (Au/V$_2$O$_5$/SiO$_2$/Si) feasible for the application as nonvolatile memory cells with nondestructive readout operation.

Due to the existence of unfilled $d$-shells, transition metals possess a set of valence states and, in compounds with oxygen, form a number of oxides. In the vanadium-oxygen system, there have been found the following phases: suboxides VO$_x$ with $x < 1$, vanadium monoxide VO, V$_2$O$_3$, the Magneli phases V$_n$O$_{2n-1}$ ($n$ = 3-9), VO$_2$, V$_6$O$_{13}$, and V$_2$O$_5$ [1, 32, 56, 57]. Lower oxides and V$_7$O$_{13}$ exhibit metallic properties, vanadium pentoxide is an insulator with the energy gap $E_g$ = 2.5 eV, and the other oxides undergo the metal-insulator transitions at different temperatures. In vanadium dioxide, for example, the MIT takes place at $T_t$ = 340 K [32]. The ability of these materials to switch their conductance between two stable states provides the basis for many electronic devices [17], including memory devices.

Figure 9 (inset) shows the device structure used in this study. Two-terminal devices have been fabricated on Si wafers, and the parameters of the samples are indicated in Table I.

First, a UV resist was deposited onto the SiO$_2$ substrate, and the 100×100 μm$^2$ regions were developed after exposure. The regimes of exposing and developing ensured almost vertical walls after etching. Next, 100 nm thick vanadium oxide film was deposited at room temperature by thermal evaporation.

**Table I.**

| Sample № | Si type | ρ Si, Ω·cm | $SiO_2$ thickness, nm/ fabrication method | $V_2O_5$ thickness, nm |
|---|---|---|---|---|
| 1 | p | 40 | 70, CVD | 100 |
| 2 | p | 40 | 30, thermal | 100 |
| 3 | n | 0.3 | 30, thermal | 200 |

Vacuum thermal evaporation was performed in a VUP-5 vacuum system at the background pressure ~ $2 \cdot 10^{-6}$ Torr using a liquid nitrogen trap. Evaporation of $V_2O_5$ powder was carried out by a flash method using a standard VUP-5 batcher. $Al_2O_3$ boats with a built-in tungsten heater were used to evaporate $V_2O_5$ powder. The distance between the boat and substrate was 10 to 12 cm, and the boat temperature was maintained at about 800°C. Before film deposition the boats were annealed at high temperature in vacuum. This technique allowed deposition of amorphous $V_2O_5$ [58]. Next, 60 nm thick gold top electrodes were deposited at room temperature using thermal evaporation. Finally, an array of $Au/V_2O_5$ 100×100 μm$^2$ junctions was fabricated on $SiO_2$/Si substrate using a lift-off technique by etching in acetone.

X-ray structural analysis has revealed the deposited $V_2O_5$ films to be amorphous – the X-ray patterns do not show any diffraction peaks [58]. In the virgin state, the junctions for all the samples demonstrate insulating properties. The electrical forming is performed by increasing the bias voltage above 50–100 V with positive or negative polarity on Au for the cases of p- and n-Si, respectively; reverse polarity does not lead to formation of structures with the memory effect. This happens as a "soft breakdown" of $V_2O_5/SiO_2$ bilayer accompanied by the formation of a switching channel with the memory effect. The I–V characteristic of sample 1 is shown in figure 9; one can see that switching of the junction resistance occurs to the LRS (6-7-8) with $R$ ~ 0.1–1 MΩ, and the resistance ratio is as high as $10^5$–$10^6$. The resistance of the HRS is ~ 100 GΩ (3-4-1-5). Too high current compliance $I_{com}$ settled at the process of electrical forming may lead to a "hard" breakdown of the $V_2O_5/SiO_2$ bilayer with nonreversible HRS-to-LRS transition.

LRS conserves after switching the bias voltage off and can be readout nondestructively applying a low probing voltage. In properly electrically formed junctions reproducible switching between LRS and HRS can be performed by continuous sweeping or pulsing the bias voltage of opposite polarity (1-2). LRS is switched back to HRS (2-3) under negative voltage around 10 V (the current of switching off is $I_{off}$). Positive voltage of about 5 V causes switching (5-6) of HRS again to LRS (the current right after switching on is $I_{on}$). LRS conserves after switching the bias voltage off and can be readout nondestructively applying a low probing voltage. In properly electrically formed junctions reproducible switching between LRS and HRS can be performed by continuous sweeping or pulsing the bias voltage of opposite polarity (1-2). LRS is switched back to HRS (2-3) under negative voltage around 10 V (the current of switching off is $I_{off}$). Positive voltage of about 5 V causes switching (5-6) of HRS again to LRS (the current right after switching on is $I_{on}$).

There is no switching if the threshold voltage is not overcome, thus the I-V characteristics have reversible character. Note also that switching from the HRS to the LRS *only occurs* when positive bias is applied to the top electrode. Similarly, switching from the LRS to the HRS only occurs when negative bias is applied to the top electrode. This means that the I-V curve shown in Figure 9 cannot be reproduced when sweeping the voltage in the opposite direction of the arrows. Based on these observations electrical switching in Au/$V_2O_5$/$SiO_2$/Si junctions must be classified as a *reversible polar effect*.

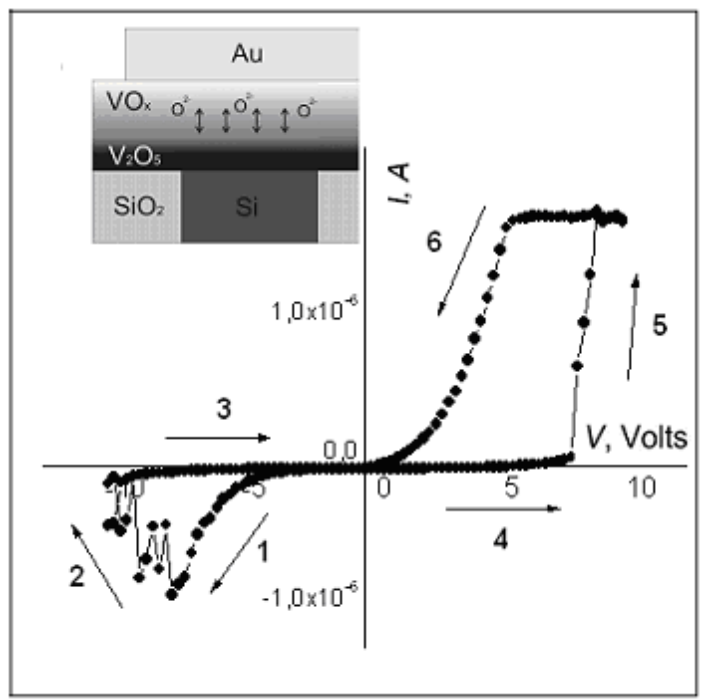

Figure 9. I-V characteristic of Au/$V_2O_5$/$SiO_2$/Si (sample 1) junctions at room temperature with arrows indicating the direction of the voltage sweep. The structure under study is also depicted schematically.

The I–V characteristic of sample 2 is qualitatively similar to the results for sample 1. The HRS-to-LRS resistance ratio for this sample is ~ $10^2$, i.e. much less than that for sample 1, because of low maximum current in LRS $I_{on}$ ~ $10^{-7}$A. The value of $I_{on}$ seems to be conditioned by the effect of current selfcompliance in LRS (6-7), allowing reversible switching of the structure without the current compliance by the sourcemeter. As shown after the study of the $SiO_2$ film breakdown, the selfcompliance effect is not linked to the $V_2O_5$ layer, and is associated with the effect on the Si/$SiO_2$ interface, which is in the depletion regime. The maximum value of the $I_{on}$ current depends on the illumination of Au/$V_2O_5$/$SiO_2$/Si junction (the Au layer is semi-transparent). At a external illumination of $10^3$ Lx, the HRS/LRS resistance ratio reaches the value of ~ $10^4$. For sample 1, the effect of illumination is weaker, and $I_{on}$ increases by 2-3 times.

The difference between samples 1 and 2 is evidently connected with the different techniques of $SiO_2$ fabrication (thermal or CVD), which, in turn, influences on the concentration of defects. The latter is important during electroforming.

The structures referred to as "sample 3" in Table I does not demonstrate any memory effects. Also, the HRS is actually absent, and the structure is in the LRS right after electroforming. The same behavior was found for the Au/V$_2$O$_5$/Si structures. Thus, the processes on the V$_2$O$_5$/SiO$_2$ interface seem to be responsible for the observed nonvolatile memory phenomena.

Qualitatively, the mechanism of reversible resistance switching seem to be associated with the electric field promoted nucleation of conducting VO$_x$ and SiO$_x$ channels in isolating V$_2$O$_5$/SiO$_2$ matrix (Figure 9). Thin SiO$_2$ layer plays the role of the insulating barrier enabling appearance of high electric field strength applied across V$_2$O$_5$ layer every time, including the initial electrical forming process, when HRS is switched to LRS. At positive bias (1-5) the transition of V$_2$O$_5$ matrix into a metallic-like state occurs through the electric field driven redistribution of oxygen ions and consequent reduction of V$_2$O$_5$ to VO$_x$ (and SiO$_2$ to SiO$_x$) in a filament perforating V$_2$O$_5$ and SiO$_2$ insulator, the oxygen depletion of the VO$_x$/SiO$_x$ interface occurs. When the structure is in the LRS, at positive bias (1-2), the motion of the oxygen ions toward the boundary of the VO$_x$/SiO$_x$ channel takes place; as a result, the conducting channel collapses. A sharp transition (2-3) from LRS to HRS can be conditioned by a positive feedback – the field in the VO$_x$/SiO$_x$ interface increases sharply at the moment of the channel collapse.

Also, we have observed the I–V characteristic of sample 1 and 2 in dynamical regime. In this case, the stable switching has been observed at frequencies up to 1 kHz. At higher frequencies, the effect of memory switching was unstable. This fact also supports a slow ionic mechanism of switching. The absence of the memory effect in sample 3 (with the *n*-type Si) might be associated with either high conductivity of silicon or high forward bias current $I_{on}$ of the SiO$_2$/Si boundary in this case.

Thus, in the present section, the results of study of a new memory element on the basis of Au/V$_2$O$_5$/SiO$_2$/Si structures have been presented. The memory effect is associated with bistable switching from a high-resistance state to a low-resistance state due to the reversible local changes of the oxygen content at the V$_2$O$_5$/SiO$_2$ interface under the action of electric field. At present, the only advantage of the structure is its high HRS-to-LRS resistance ratio. The study of its time characteristics is going on. Finally, the results presented may be useful for potential applications in memory devices.

### 3.2. Unipolar Resistive Memory Switching in Niobium and Nickel Oxides

In this section we present the results on the switching effect in thin-film MOM (metal-oxide-metal) structures based on niobium oxide. The structures under study were obtained by oxidation of the metal surface (either Nb foils or layers deposited on glass-ceramic or silicon substrate) with subsequent deposition onto the surface of the oxide film of Al- or Au-electrodes. Oxide films were prepared by anodic oxidation of niobium in 0.1 N aqueous solution of phosphoric acid (H$_3$PO$_4$) at room temperature in the galvanostatic regime. For different samples, the oxide film thickness was in the range from 100 to 300 nm.

For the measurements of the electrical properties, we used a two-probe system on the basis of a Keithley Model 2410 SourceMeter. The system was designed to measure current-voltage characteristics of micro- and nanostructures both in galvanostatic and in potentiostatic modes. A ramp voltage up to 15 V of a certain polarity was applied to the MOM structure, and the current was limited in order to prevent an irreversible breakdown of the structure.

Figure 10 shows a typical current-voltage dependence of the Nb-oxide based structures after electrical forming. The obtained I-V characteristics demonstrate *unipolar* resistive switching.

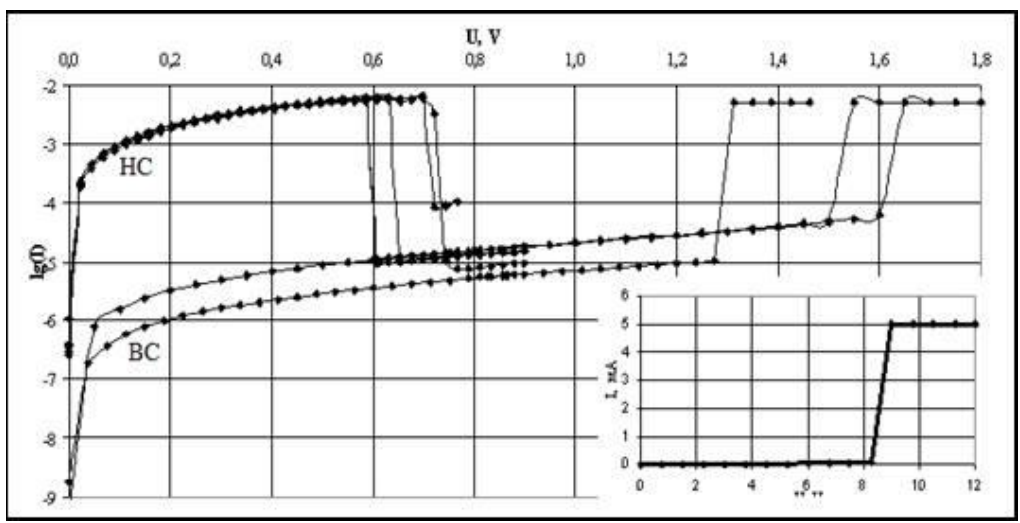

Figure 10. Current-voltage characteristics of the MOM structure based on Nb oxide. The inset shows the I-V curve during EF.

In case of unipolar switching, the conductivity change does depend on the amplitude of the applied voltage, and the polarity is not important. When initial structure under an electric field transits into the LRS, this process also is referred to as the process of EF, likewise this takes place in CGS [15-17, 48].

After forming, the formed memory cells could be reversible switched back to, upon HRL application of the threshold voltage (reset process). Switching from HRS to LRS (the recording process) is achieved by using a larger threshold voltage than the voltage reset. Resistance ratio $R_{HPS} / R_{LRS}$ is as high as $10^2$-$10^3$. In the process of recording, the current should be limited.

The physical mechanism of the effect of bistable switching in unipolar structures based on Nb anodic oxide requires further study, based on a more experimental data, and analysis of different models offered in the literature [59-61]. Note that in the amorphous oxide, $Nb_2O_5$, there is also monostable (threshold) [61] and bistable (memory effect) with *bipolar* switching [60]. What type of switching (threshold or memory, bi- or mono-polar) is implemented in a real MOM structure – it depends on the conditions of the FE process [15] and current compliance during the measurements.

In addition to the considered here oxides of V and Nb, as was already said above, a number of other TMOs demonstrate the memory switching effects [12-18, 26, 38-42, 45, 60-63]. Therefore, finally, we present some experimental results (see Figures 11 and 12) on bistable unipolar switching in NiO [26, 64-66]. Note that nickel oxide seems to be the most investigated material for ReRAM applicatioins, at least, experimentally [12, 26, 67-71].

The typical current-voltage characteristics of the MOM structure on the basis of magnetron sputtered NiO with Pt top and bottom electrodes, which were used for modeling, presented in the next Section, are shown in Figures 11 and 12. Also, Figure 11 demonstrates also a cross section of the structure, giving geometry and some indications.

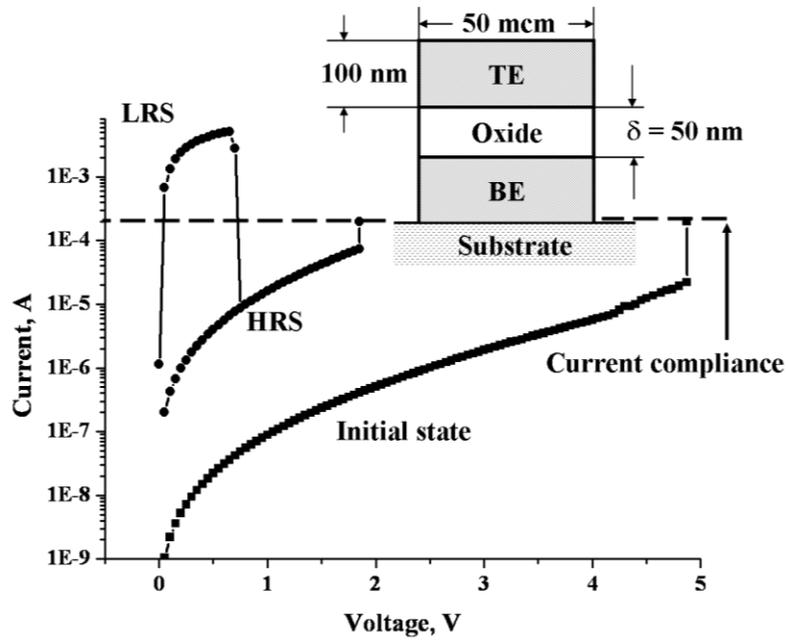

Figure 11. Typical current-voltage characteristic for Pt-NiO-Pt structure with nonvolatile unipolar switching in voltage controlled regime of the measurement. TE and BE are Pt top and bottom electrodes.

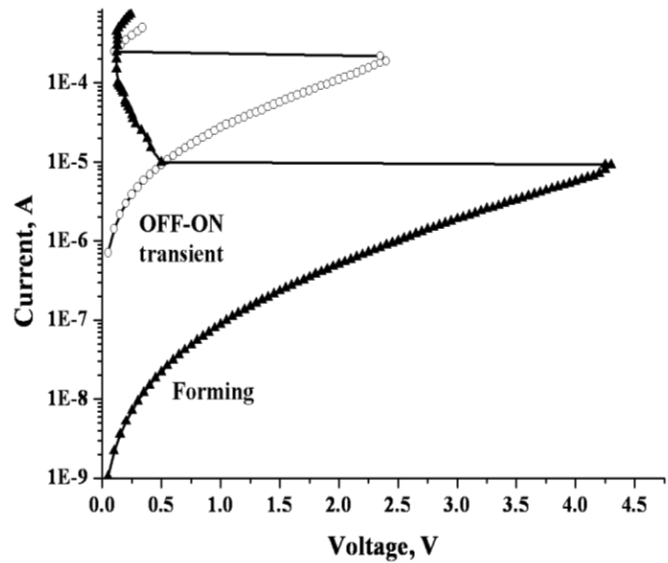

Figure 12. Current-voltage characteristic for forming and OFF-ON transient in current controlled regime of the measurement.

## 4. MODEL OF CONDUCTING CHANNEL FORMATION INSIDE THE DIELECTRIC OXIDE MATRIX

In this section we present a model of the electrically actuated formation of the nanosized metal filament inside an oxide matrix. The dielectric breakdown of the oxide with conditions of the adequate current compliance and subsequent capacitance discharge of the energy, which have been stored in thin film oxide capacitor structure before breakdown, results in sharp local temperature growth and, as a result, in fast local oxide reduction. The so-called Soret state with metal segregation on the center of the high temperature region is established by temperature gradient-driven diffusion. The nanosized metal filament is quenched by fast temperature drop after capacitance discharge ending. At the next biasing, the local domain with high resistance and high electric field is created near the cathode end of a filament by metal electromigration due to the electron wind induced by high density electron current. A part of the metal filament is transformed in oxide by the subsequent fast electric field enhanced thermal oxidation.

The study was initiated by the experimental observation of the electrically actuated nonvolatile switching between two resistance states in thin films oxide structures [64]. To date this phenomenon is considered as promising candidate for development of the high density stackable nonvolatile memory. The nonvolatile switching usually can be divided into two different groups - polarity dependent switching and practically symmetrical unipolar switching (see Figure 13). There exists a common understanding of the bipolar switching with memory. The majority of the researches show that bipolar switching is interface phenomenon in which the interface properties (interface transition layer resistance [65] or Shottky barrier height and form [66]) are modified by high field ionic transport. On the other hand, a mechanism of the unipolar switching with memory is far from understanding.

As was said above, the most investigated oxide, at least experimentally, is NiO [67-71]. The experimental results show that first any polarity electrical biasing of the initial oxide structure with semiconductor type of the conductivity induces the growth of the thin filament (or filaments) with metal conductivity inside oxide matrix (forming).

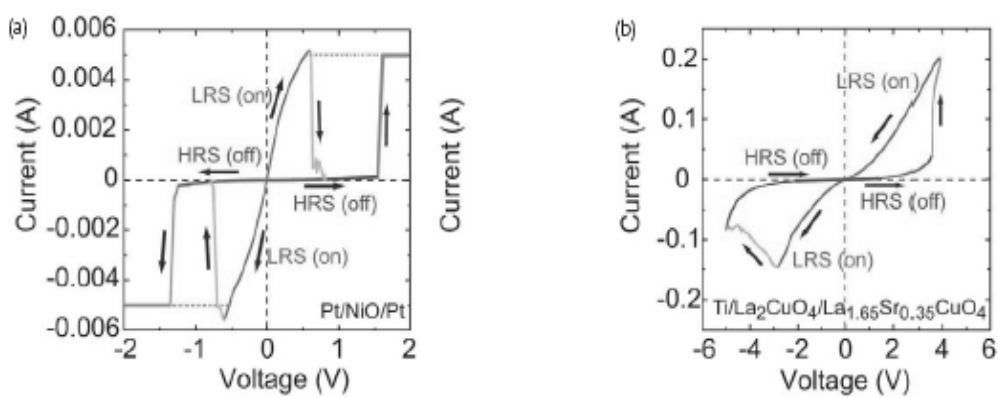

Figure 13. (a) I-V characteristics of unipolar switching in Pt/NiO/Pt structure and (b) bipolar switching in Ti/$La_2CuO_4$/$La_xSr_yCuO_4$ structure [12].

The second any polarity electrical biasing can ruptures metal filament that returns semiconductor properties to structure. Further, a transition between HRS and LRS) can be repeated many times.

The I-V curve of the initial oxide structure (Figure 12) during first polarization measured at current controlled regime shows that forming can be classified as irreversible threshold switching with unstable section of the current controlled negative differential resistance (NDR). These features of the forming allow considering it as hard breakdown of the insulator oxide. Note that followed LRS-HRS transition can be observed only if adequate compliance current, $I_C$, which has been used. The electrical bias with high magnitude of the current compliance transfers structure into LRS state which can not be ruptured by next voltage input. The lowest level of the compliance current is defined by the value of the breakdown threshold current.

There is no a universal mechanism of thin films breakdown, but all researches allocate two process stages. At the first stage a sudden reduction of the insulator resistance driven by electronic or electrothermal positive feedback mechanism occurs. The NDR region appears in the I-V curve, and a thin conductive channel is formed between electrodes. During second stage of the breakdown the permanent conductive filament, whose structure and chemical composition differs from native oxide, is formed inside the insulator [72]. Taking into attention this universal phenomenological behavior of the thin film insulators, the first breakdown stage is not so important for presented model. When any electronic or electrothermal instability has been initiated and, as a result, the conductive channel has been formed, the temperature increases due to the Joule heating of this local region which could result in a local thermochemical modification of the oxide.

There are two approaches to the estimation of the energy dissipation region size. The universal thermodynamic consideration [73] shows that in system with initial uniform current distribution the trend of the current to collect in local domain is governed by the principal of least entropy production. Using approximation which have been developed in [73], we calculate that radius $a$ of cross-sectional area of the cylinder conductive channel, which have been formed after first breakdown stage, is 5 nm.

Another approach is based on a strong nonuniform distribution of the current in pre-breakdown state of the defect insulator [74, 75]. The statistical model of the electric field enhancement by local geometric thinning of the oxide thickness assumes that the size of the high conductive path after first stage of the breakdown is the same as interface irregularities size – 5 nm [74]. Note also that the other high conductive defect in polycrystalline NiO is the grain boundaries which size have been measured as 5-10 nm [74, 76]. Therefore, the 5 nm as the dimension scale for $a$ is a reasonable estimation.

It is obvious that there are two energy sources for Joule heating of the conductive domain. At first, it is necessary to take into account the action of the direct current trough conductive channel. The density of the dissipated power can be calculated as $P_{DC} = I_c \times V_c / v$, where $I_c$ is the current compliance, $V_c$ is the voltage which corresponds to $I_c$, and $v$ – the volume of the conductive domain. Obviously that $V_c \leq V_F$, where $V_F$ is forming voltage, because, right after the first breakdown stage, the structure has I-V charactistic with current-controlled NDR. Accepting a high current domain as cylindrical body with basis radius $a = 5$ nm and oxide thickness as height $\delta = 50$ nm we obtain $P_{DC} = 5 \times 10^{13}$ W×cm$^{-3}$.

Before breakdown, the structure is the capacitor with capacitance of $C$ which is charged up to the voltage of $V_F$. At the second breakdown stage, this energy is liberated by electrical discharge through conductive channel. The storage energy can be written as $E_C = [C \times (V_F - V_C)^2]/2$ and it is equal to $10^{-13}$ J for the analyzed sample. The capacitance discharge power density $P_C$ changes during energy liberation process but we assume that capacitance discharges occurs with constant rate at characteristic time $\tau_0 = C \cdot V_c/I_c = 10^{-9}$s, and under such an assumption the power density $P_C = E_C/(v \times \tau_0) \approx 10^{16}$ W×cm$^{-3}$, therefore $P_C \gg P_{DC}$. Note that $\tau \leq 10^{-9}$s and it is typical transient time of second breakdown stage for many thin insulator films [72, 75].

For estimations of the temperature space-time distributions we will assume that the heat production is confined to a conductive cylinder with height $\delta$ and radius $a$. The temperature will be determined by oxide thermal conduction in both axial and radial direction, and spreading thermal resistance in electrodes. When $a$ is enough small as compared to other dimensions of the structure (thickness of oxide, thickness and size of metal electrodes) we can assume that the temperatures of electrodes and oxide matrix volume are equal ambient. Also, for thin conductive path, the cylinder lateral surface is much greater than the basis surface and we can assume that the radial heat flow will dominate.

The steady-state solution of the heat equation in cylindrical coordinates with heating by a line source of strength $Q_C = P_C/\delta$ along cylinder axis and with $T = 0$ at $r = d$, and $dT/dr = 0$ at $r = 0$ is [77]

$$\Delta T_C = \frac{Q_C}{2\pi K_{NiO}} \ln\left(\frac{d}{r}\right), \tag{5}$$

In practice $d$ will not be the sample or electrode size, and in order to obtain a realistic estimation we therefore replace $d$ with the oxide thickness $\delta$. Taking nickel oxide thermal conductivity $K_{NiO} = 0.71$ W/(cm$^0$C), we obtain the rise of temperature on the conductive channel boundary $(r = a)$ $\Delta T_C \approx 4000^0$C that is more than the oxide melting point, $T_{mNiO} = 1990^0$C.

At last stage of the forming process, when capacitance discharge will be finished, the temperature of the hot region in oxide matrix for the time $\tau_T \leq 10^{-10}$ should drop down to values which would be defined by heat production due to current compliance. For estimation of this temperature we can use Equation (5) with the same problem geometry but replacing $P_{DC}$ with $P_C$, which yields $\Delta T_{DC} \leq 100^0$C.

The melting time, $t_m$, can be evaluated from quasistationary approximation of the Stefan problem of cylindrical body melting due to a line heat source of strength $Q_C$ at $r = 0$. The appropriate solution is given by the equation [78]: $t_m = \pi a^2 \gamma_{NiO} L_{fNiO}/Q_C$, where $L_{fNiO}$ is the oxide latent heat of fusion. Taking $L_f = 0.78$ kJ/g, $t_m = 10^{-13}$s. We have thus arrive to a conclusion that, during high temperature forming stage, the NiO conductive domain and some region around it should be transformed to melting state.

Second process which should be considered at high temperature forming stage is oxide reduction. Extensive studies of the NiO reduction have appeared in literature and the important result in frames of our consideration is that the reduction of NiO is irreversible, since the equilibrium constant $K_{eq}$ of the reduction reaction reaches $10^3$ in high temperature limit [79]. Note that oxide reduction due to direct thermal decomposition is reaction-limited process and we can neglect diffusion of the reaction products for estimation of the reduction time scale. Consider NiO reduction as first-order reaction with respect to Ni we can write solution of the reaction kinetic equation as: $C_{Ni}/C_{NiO} = [1-exp(-kt)]$, where $C_{Ni}$ is the Ni concentration, $C_{0NiO}$ is the initial NiO concentration, and $k = k_0\, exp(-E_{rmol}/RT)$, where $k$ is the reduction reaction rate constant, $E_{rmol}$ is the molar activation energy and $R$ is the gas constant. Using experimental values: $E_{rmol} = 90$ kJ/mol and $k_0 = 6\times10^{13}$ s$^{-1}$ [80] we can estimate the characteristic time constant of the NiO reduction as $\tau_R = 1/k < 10^{-11}$s. We have to conclude that NiO melting region and nearest solid state region with sufficiently high temperature must be converted to mixture of the Ni and O atoms during capacitance discharge regime.

The presence of the strong temperature gradients can result in temperature gradient-driven diffusion (thermomigration) [81]. Thermomigration in solid is small and therefore it can be usually neglected as compared to concentration diffusion. In a heat flow transient induced by electrical discharge, however, temperature gradient is of the order of $10^8$ $^0$C/cm and thermal diffusion contribution cannot be excluded, especially in the melt state of the oxide. If a homogeneous binary compound is placed in a temperature gradient, a redistribution of the constituents can occur, and one constituent migrates to the cold end of the specimen and other – to the hot end. This phenomenon is called the Soret effect [82]. The direction of the migration and values of the mass flows are defined by the transport heat $f$ of the diffusing ions $Q^*$. The values of the $Q^*$'s for Ni and O thermomigration in NiO are unknown. However, we can use the approaches which were developed for liquid conductive compounds [83]. Indeed, in this theory assuming that the liquid is a dense gas and applying the thermo-transport theory in binary gas mixtures, the direction of the diffusion is determined primarily by the mass differences: the lighter component migrates to the warmer end and the heavy component to the cold one. Taking this fact into account, we can assume that Ni ions migrate towards the hot region, whereas the O ions diffuse to periphery of the melt region. As a consequence, a temperature gradient drives the establishment of concentration gradients. In the stationary state this concentration gradient depends on the boundary conditions. As melt region are closed for the exchange of oxygen with the surrounding gas phase, the process ends up with zero atom fluxes, defining the so-called Soret state with Ni rich region in the center of the melt.

The data given in Figure 14 confirm an opportunity of an establishment of the Soret state at high temperature stage of the forming. The presented results are the SIMS images of the O and Ni distribution near NiO-Pt interfaces for initial oxide structure and after forming. We can see that only O diffuses away from local nonhomogeneous regions of the NiO during forming. Assuming that these local regions have highest conductivity and, as consequence, high temperature due to Joule heating, the atoms redistribution can be defined by thermomigration and Soret state establishment.

At the last forming stage, when liberation of the capacitance energy will be finished, the temperature drops down to above estimated low values due to the thermal conductivity, and the solidification of the melting region should occur in time $t_s$. This time can be estimated from time dependence of the solidification front position $R(t)$. In our case temperature difference between solid and liquid phases near the interface is not so big and we can assume that liquid has melting temperature and temperature profile in the solid is linear. The solution of the appropriate Stefan problem can be written as [77, 78]: $t_s = a^2\gamma_{Ni}L_{fNi}/2K_{Ni}T_m$. The value of $t_s$ is less than $10^{-11}$s and fast solidification should quench the Ni filament inside the oxide matrix.

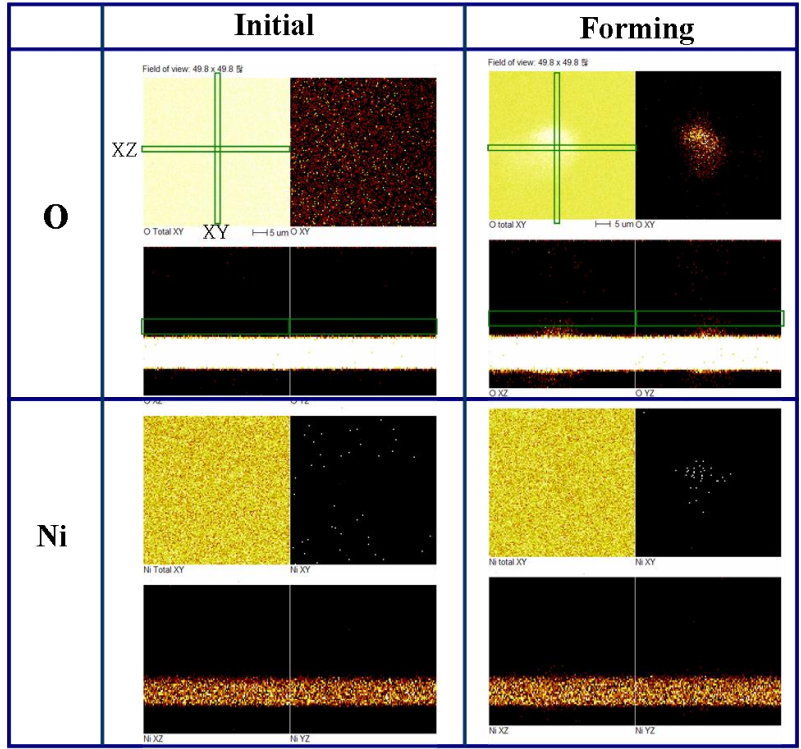

Figure 14. SIMS images of the Ni and O distributions near NiO-Pt interfaces in initial state and after forming.

The low value of the diffusion coefficient for Ni diffusion in NiO and electrode materials (during the final low-temperature stage of electroforming [84]) allows assuming that the influence of Ni diffusion on final Ni filament size is negligible. The oxidation process at the Ni-NiO interface could also be rather slight, because, for this reaction at low temperatures, the oxidation rate is limited by slow oxygen diffusion transport toward the NiO-Ni interface [80].

The strict solution of the problem of the Ni filament size, $R_f$, is based on considering of the energy conservation equation, but the simple estimations show that heating and heat transfer terms are much less in comparison with melting and chemical reaction terms. Assuming that the volume of the melt is $v=\pi R_f^2 \delta$ and that intensive thermal reduction is going only in molten region, we can write more simple integral energy conservation equation for steady-state regime

$$E_C = Q_{melting} + Q_{reduction} = \gamma_{NiO} v L_{fNiO} + \gamma_{NiO} v \frac{E_{Rmol}}{M_{mol}}, \qquad (6)$$

where $M_{mol}$ is NiO molar mass. The solution of Eq.(6) yields

$$R_f = \sqrt{\frac{E_C}{\pi \delta \gamma_{NiO} (L_{fNiO} + \frac{E_{Rmol}}{M_{mol}})}} \tag{7}$$

and we obtain $R_f \approx 7$ nm.

Thus, we can conclude that the Ni melt filament with radius $R_f$ is formed inside NiO during energy liberation stage. After discharge the fast solidification of the Ni melt should occur which provides a stable metallic LRS of the oxide structure after the voltage turned off.

A similar model has been proposed recently in the work [26] where it has been shown that the transition from insulating to metallic conductivity in NiO first results from purely electronic threshold switching, which then causes the formation of a conducting filament by the local high current and high temperature conditions. A set transition time below 1 ns has been evidenced, and the impact of parasitic capacitance has been confirmed by numerical simulations of threshold switching and Joule heating [26]. Also, the $TiO_2$ based sandwich structure studied in [85] has demonstrated behavior resembling the above described processes, i.e. electroreduction and drift process triggered by high electric fields and enhanced by Joule heating [85]. Also, in this work, the results are reported revealing stable rectification and resistive-switching properties of a $Ti/TiO_2/Pt$ structure. The oxygen migration and localized conductive filaments play important roles in not only the resistive-switching of ReRAM, but also in the process of the rectification of oxide diodes. The rectification properties stable up to 125°C and $10^3$ cycles under about 3 V sweep without interference with resistive-switching. This shows a satisfactory reliability of $TiO_2$ MIM diodes for future 1D1R (one diode – one resistor) ReRAM applications [85].

The reverse process of the Ni filament interruption, and the transition of the structure from LRS back to HRS, is rather more complicated for calculations [86], and we will not develop here all these calculations and estimates. It should be noted, however, that all the above described phenomena – electro- and thermo-diffusion, the Soret effect, electronic wind and so on – play an important role, although, all of them are different in their intensity and, thereby, in relative contribution to the mechanism of the LRS – HRS transition. In other words, several processes are involved during this LRS – HRS transition, but their importance (in order to interrupt the Ni-metal filament) will be defined by a parity of their time scales to capacitance discharge time. Evidently, we should consider melting, thermoreduction of oxide, reoxidation, diffusion and solidification of the components of the reduction-oxidation reaction. Part of these processes will be going in parallel and interdependently, but their parity can be defined by separate consideration of temporary evolution of each process. In more detail, all these effects have been considered earlier in the work [86].

## CONCLUSION

Transition-metal oxide films, demonstrating the effects of both threshold and nonvolatile memory resistive switching, have been proposed as candidate materials for storage-class memory [11, 12]. In this work we have described experimental results on threshold switching in a number of diverse transition metal oxides (V, Ti, Fe, Nb, Mo, W, Hf, Zr, Mn, Y, and Ta), and the amorphous oxide films of these metals have been obtained by anodic oxidation.

In the case of nonvolatile resistive switching, on the basis of *I–V* characteristics, the switching behaviors can be classified into two types: unipolar (nonpolar) and bipolar. Conceptually unsophisticated design of such a MOM memory cell allows easily scalable cross-point ReRAM architecture with nanometer cell dimensions. This, in turn, provides the basis for 3D integrated terabit memory with a multi-layer stackable structure [13].

Next we have studied the effects of bistable resistive switching in MOM and MOS structures on the basis of transition metal oxides. Oxide films were prepared by electrochemical oxidation of V and Nb, as well as by vacuum evaporation of vanadium pentoxide and NiO onto Si-SiO$_2$ and metal (in particular, Pt) substrates. Possible mechanisms of bistable switching, including a compositionally-induced *metal-insulator transition*, are proposed to account for the memory properties of the "Si/vanadium oxide/metal" and "Nb/niobium oxide/metal" structures.

In conclusion, oxide materials and physical-chemical phenomena therein, termed as "*oxide electronics*" [3-9], is a promising direction for alternative electronics beyond silicon, along with (and in addition to) such up-to-date approaches as molecular electronics, spintronics or superconducting electronics [4]. That is why we postulate that, in spite the fact that TMOs and their properties had being studied for many years, this new breakthrough to oxide electronics does give them a novel hypostasis. On the other hand, Si-VO$_x$ structures may serve as an example demonstrating the availability of hybrid devices based on both traditional silicon technology and new memory technologies. It should be noted that the class of oxide materials potentially suitable for ReRAM applications is not limited by the above mentioned materials: see, e.g., Table 1 in the recent survey [18], representing more than 40 different oxides and systems. Really ample quantity to select! All the aforesaid gives hope of forthcoming oxide electronics utilizing metal-insulator transitions [87] and other phenomena in TMOs enabling and accelerating advances in information processing and storage beyond conventional CMOS scaling technology.

## ACKNOWLEDGEMENTS

This work was supported by the Russian Federation Ministry of Education and Science within the "Scientific and Educational Community of Innovation Russia" Programme (2009-2113), through contracts no. 02.740.11.0395, 02.740.11.5179, 14.740.11.0137, P1156, P1220, and "Development of Scientific Potential of High School (2009-2011)" Programme, project no. 12871.